\documentclass[aps,prb,twocolumn,amsmath,amsart,amssymb,superscriptaddress]{revtex4-2}
\usepackage[colorlinks=true, linkcolor=blue, citecolor=blue, urlcolor=blue]{hyperref}

\usepackage{graphicx}
\usepackage[T1]{fontenc}
\usepackage{latexsym}
\usepackage{amsmath}
\usepackage{lmodern}
\usepackage{color}

\usepackage[normalem]{ulem}
\usepackage{times}
\usepackage{txfonts}
\usepackage{soul}
\usepackage{xcolor}

\newcommand{\CVS}{CsV$_3$Sb$_5$}

\begin{document}

\title{$c$-axis strain tuning of superconductivity and symmetric elastoresistivity in CsV$_3$Sb$_5$}

\author{Xiaoran Yang\textsuperscript{\S}}
\affiliation{Center for Advanced Quantum Studies, School of Physics and Astronomy, and Key Laboratory of Multiscale Spin Physics (Ministry of Education), Beijing Normal University, Beijing 100875, China}
\affiliation{Center for Correlated Matter and School of Physics, Zhejiang University, Hangzhou 310058, China}
\author{Yutong Li\textsuperscript{\S}}
\author{Chunyi Li}
\author{Qi Tang}
\email{17396266220@163.com}
\affiliation{Center for Advanced Quantum Studies, School of Physics and Astronomy, and Key Laboratory of Multiscale Spin Physics (Ministry of Education), Beijing Normal University, Beijing 100875, China}

\author{Jiawen Zhang}
\affiliation{Center for Correlated Matter and School of Physics, Zhejiang University, Hangzhou 310058, China}
\affiliation{State Key Laboratory of Silicon and Advanced Semiconductor Materials, Zhejiang University, Hangzhou 310058, China}
\author{Yu Song}
\affiliation{Center for Correlated Matter and School of Physics, Zhejiang University, Hangzhou 310058, China}
\author{Huiqiu Yuan}
\affiliation{Center for Correlated Matter and School of Physics, Zhejiang University, Hangzhou 310058, China}
\affiliation{State Key Laboratory of Silicon and Advanced Semiconductor Materials, Zhejiang University, Hangzhou 310058, China}

\author{Xingye Lu}
\email{luxy@bnu.edu.cn}
\affiliation{Center for Advanced Quantum Studies, School of Physics and Astronomy, and Key Laboratory of Multiscale Spin Physics (Ministry of Education), Beijing Normal University, Beijing 100875, China}

\begin{abstract}
The kagome metal CsV$_{3}$Sb$_{5}$ hosts an intriguing interplay between charge-density-wave (CDW) order and superconductivity that is highly sensitive to lattice distortions. However, determining the specific roles of the in-plane ($A_{1g,1}$) and out-of-plane ($A_{1g,2}$) symmetric strain channels has been hindered by their intrinsic mixing in conventional piezo-based experiments. Here, we combine in-plane uniaxial strain with direct $c$-axis compression to independently access and disentangle these symmetry-resolved responses in CsV$_{3}$Sb$_{5}$. We reveal that $c$-axis compression drives a massive, linear enhancement of the superconducting transition temperature ($T_c$) alongside a suppression of $T_{\rm CDW}$. The tuning efficiency of this out-of-plane deformation acts with an opposite sign and far exceeds that of in-plane strain, demonstrating that $c$-axis lattice control dictates the phase competition. Furthermore, by isolating the pure elastoresistivity coefficients, we find that the out-of-plane cross-coupling coefficient ($m_{13}$) is comparable in magnitude but opposite in sign to the in-plane response ($m_{11}+m_{12}$). Unlike the sharply peaked in-plane response, $m_{13}$ exhibits a distinct, order-parameter-like onset across the CDW transition. Our results establish that out-of-plane lattice control plays a dominant role in tuning the intertwined states in CsV$_{3}$Sb$_{5}$ and provide a general pathway for resolving strain-coupled electronic responses in layered quantum materials.

\end{abstract}

\maketitle

\section{Introduction}
Kagome metals have emerged as a particularly rich platform for exploring correlated and topological electronic states, owing to their characteristic lattice geometry, which hosts Dirac crossings, flat bands, and van Hove singularities \cite{rmp2026,yin2022topological,wang2023quantum}. Among these, the $A$V$_3$Sb$_5$ family ($A$ = K, Rb, Cs) has attracted substantial interest due to the simultaneous presence of nontrivial band topology \cite{kang2022two,hu2022topological}, charge-density-wave (CDW) order \cite{PhysRevB.105.195136,Li2022conjoinedCDW}, and superconductivity \cite{ortiz2020}, making them ideal for studying intertwined electronic instabilities \cite{ortiz2019new,ortiz2020,kang2022two,zhao2021cascade,xie2022electron,subires2023order,wang2025soft,mcguinness2025soft}. CsV$_3$Sb$_5$, in particular, has become a focal system for investigating the interplay between lattice \cite{PhysRevB.105.195136,mcguinness2025soft}, electronic correlations \cite{zhao2021cascade,zheng2022emergent}, and symmetry-breaking phenomena \cite{Xiang2021two-foldCVS,nie2022charge,Li2022conjoinedCDW}, as well as for exploring strain-tunable electronic responses \cite{qian2021revealing,liu2024absence,asaba2024evidence,frachet2024colossal,yang2023thekagome}.

CsV$_3$Sb$_5$ crystallizes in the hexagonal $P6/mmm$ structure, composed of V–Sb kagome layers separated by Cs atoms. Within the kagome layers, the V sublattice forms a two-dimensional corner-sharing network coordinated by Sb octahedra [Fig. 1(a)] \cite{ortiz2019new}. The system undergoes a CDW transition at $T_{\mathrm{CDW}}\approx 94$ K, followed by a superconducting transition at $T_c\approx 3$ K \cite{ortiz2020}. The CDW order has been well characterized experimentally and is believed to originate from a combination of electronic instabilities near van Hove singularities and electron–phonon coupling \cite{kang2022two,mcguinness2025soft}. Recent studies have also reported evidence of rotational-symmetry breaking within the CDW phase \cite{nie2022charge}, pointing to an unconventional and potentially intertwined electronic state that may couple to lattice distortions and superconductivity.

Understanding the interplay between superconductivity and the CDW in CsV$_3$Sb$_5$ is essential for elucidating the microscopic mechanisms underlying its low-temperature electronic ground state. Uniaxial stress and strain provide a symmetry-sensitive tuning parameter to probe this interplay \cite{qian2021revealing,yang2023thekagome}. Previous studies have shown that $a$-axis strain shifts $T_c$ and $T_{\rm CDW}$ in opposite directions, indicative of strong competition, and comparison with hydrostatic pressure suggests that the dominant tuning may arise from the accompanying change of the $c$-axis lattice constant rather than explicit in-plane $C_6$ symmetry breaking \cite{qian2021revealing}. More recent measurements, with strain applied along the $a$ axis and its in-plane perpendicular direction, found nearly identical $dT_c/d\varepsilon$ and $dT_{\rm CDW}/d\varepsilon$, indicating that the response is dominated by the fully symmetric $A_{1g}$ channel rather than by an in-plane anisotropic component \cite{yang2023thekagome}. While these observations are consistent with an important role of Poisson-induced $c$-axis strain, they do not disentangle the in-plane $A_{1g,1}$ and out-of-plane $A_{1g,2}$ contributions within the $D_{6h}$ point group. Thermodynamic measurements have likewise pointed to a notable sensitivity to $c$-axis stress \cite{frachet2024colossal}, but direct experimental tuning of the intertwined CDW and superconducting states by controlled $\varepsilon_{zz}$ remains comparatively unexplored.

\begin{figure*}
\includegraphics[width=16cm]{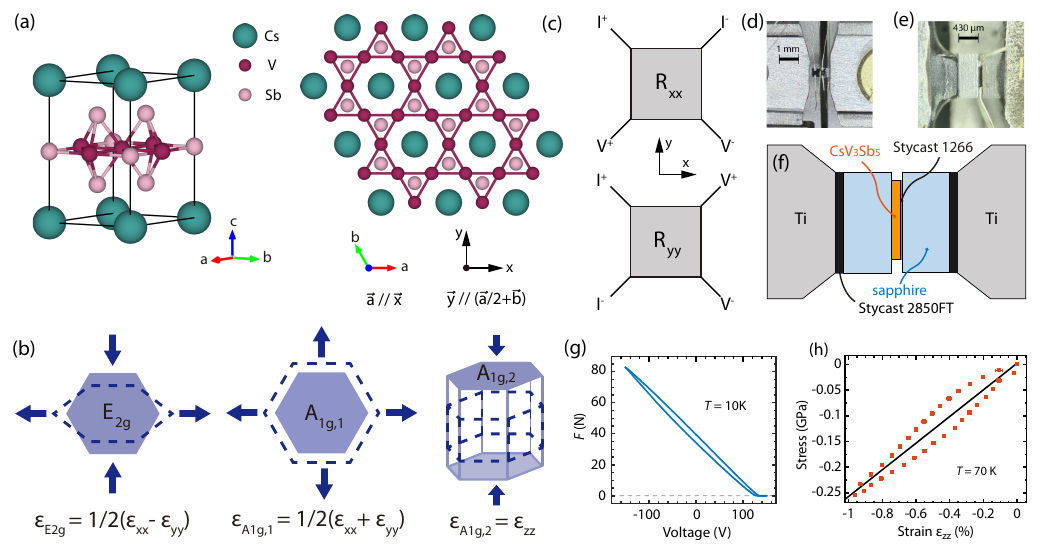}
\caption{Crystal structure, strain symmetries, and experimental configurations for in-plane and $c$-axis measurements in {\CVS}.
(a) Crystal structure of {\CVS} and top view of the V–Sb kagome layer, with crystallographic axes indicated.
(b) Schematic illustration of the symmetry-resolved strain components in the hexagonal $D_{6h}$ point group: $E_{2g}$ anisotropic in-plane strain $\varepsilon_{E_{2g}}=\tfrac{1}{2}(\varepsilon_{xx}-\varepsilon_{yy})$, in-plane symmetric strain $\varepsilon_{A_{1g,1}}=\tfrac{1}{2}(\varepsilon_{xx}+\varepsilon_{yy})$, and out-of-plane symmetric strain $\varepsilon_{A_{1g,2}}=\varepsilon_{zz}$.
(c) Modified Montgomery configuration used to simultaneously determine the two in-plane resistivity components $\rho_{xx}$ and $\rho_{yy}$ (and their strain derivatives), with $x\parallel \mathbf{a}$ and $y\parallel(\tfrac{1}{2}\mathbf{a}+\mathbf{b})$.
(d) Photograph of a representative sample mounted for in-plane uniaxial strain measurements on the bowtie-shaped titanium platform, with four electrical contacts arranged for the modified Montgomery method.
(e) Photograph of a representative sample mounted for $c$-axis compression measurements; the platelet thickness along $c$ is $\sim 90~\mu$m.
(f) Cross-sectional schematic of the $c$-axis compression geometry. A {\CVS} crystal is bonded to a sapphire plate using a thin Stycast 1266 layer and compressed between two titanium anvils through sapphire and epoxy (Stycast 2850FT/1266) interfaces.
(g) Force $F$ recorded by the capacitive sensor as a function of FC100 drive voltage at $T=10$ K. A voltage window with essentially constant capacitance (hence constant $F\approx 0$) is used to identify the zero-force point after cooldown.
(h) Stress–strain relation under $c$-axis loading at $T=100$ K, where the applied stress is obtained from the calibrated force and sample area, and $\varepsilon_{zz}$ is determined independently by cryogenic digital image correlation; the solid line indicates a representative linear fit used to extract the apparent Young’s modulus of the mounted sample configuration.
}
\label{fig1}
\end{figure*}

Meanwhile, the possible emergence of electronic nematicity intertwined with the CDW phase of CsV$_3$Sb$_5$ has stimulated intense interest \cite{nie2022charge}. Elastoresistance has served as a particularly sensitive, symmetry-resolved probe of this putative nematic susceptibility \cite{kuo2013measurement}. Early elastoresistance measurements highlighted an enhanced in-plane symmetry-breaking response in $E_{2g}$ channel $\zeta_{E_{2g}} \equiv \tfrac{1}{2} \left( \frac{\Delta\rho_{xx}}{\rho_{xx}} - \frac{\Delta\rho_{yy}}{\rho_{yy}} \right)$ with coefficient $m_{E_{2g}}=m_{11}-m_{12}$, which were interpreted in terms of rotational-symmetry breaking within the CDW state \cite{nie2022charge}. Subsequent elastoresistivity and thermodynamic studies, however, emphasized instead that the dominant strain response resides in the fully symmetric ($A_{1g}$) sector \cite{liu2024absence,frachet2024colossal,asaba2024evidence}.
In the hexagonal $D_{6h}$ point group, the fully symmetric strain sector decomposes into an in-plane isotropic component, $\varepsilon_{A_{1g,1}} = \tfrac{1}{2}(\varepsilon_{xx} + \varepsilon_{yy})$, and an out-of-plane component, $\varepsilon_{A_{1g,2}} = \varepsilon_{zz}$ [Fig. 1(b)]. 
In commonly piezo-based in-plane strain experiments, the applied $\varepsilon_{xx}$ 
induces an $\varepsilon_{yy}=-\nu_{12}\varepsilon_{xx}$ and an $\varepsilon_{zz}=-\nu_{13}\varepsilon_{xx}$ via Poisson effect. Thus, the reported effective $A_{1g}$ elastoresistivity $\zeta_{A_{1g}} \equiv \tfrac{1}{2} \left( \frac{\Delta\rho_{xx}}{\rho_{xx}} + \frac{\Delta\rho_{yy}}{\rho_{yy}} \right)$ ($m_{A_{1g}}$) generally contains both in-plane and $c$-axis contributions: $\zeta_{A_{1g}} = (m_{11} + m_{12}) \,\varepsilon_{A_{1g,1}} + m_{13} \,\varepsilon_{zz}$ with corresponding coefficient expressed as:  
\begin{equation}
m_{A_{1g}} = m_{11} + m_{12} - \frac{2\nu_{13}^{\text{s}}}{1 - \nu_{12}^{\text{p}}} m_{13},
\label{eq1}
\end{equation}
where $\nu_{13}^{\text{s}}$ and $\nu_{12}^{\text{p}}$ denote the poisson ratio of {\CVS} and the platform (piezoelectric stack here) used for measurements, respectively \cite{liu2024absence,frachet2024colossal,asaba2024evidence}.
Given that uniaxial strain and pressure studies of $T_c$ and $T_{\rm CDW}$ emphasize the outsized influence of the $c$-axis lattice, separating the in-plane $A_{1g,1}$ and out-of-plane $A_{1g,2}$ contributions becomes essential for interpreting the large symmetric elastoresistivity in CsV$_3$Sb$_5$ and for identifying which fully symmetric strain component couples most strongly to the intertwined CDW and superconducting states.

In this work, we address this outstanding symmetry-resolution problem by combining in-plane uniaxial strain with direct $c$-axis compression of CsV$_3$Sb$_5$ single crystals using a uniaxial strain cell, with strains precisely calibrated via cryogenic digital image correlation \cite{mo2024cryogenic}. By independently controlling the in-plane and out-of-plane components of the fully symmetric strain sector, we extract the separate contributions of $A_{1g,1}$ and $A_{1g,2}$  to the elastoresistive response and disentangle the corresponding strain derivatives of the superconducting transition. We find that $c$-axis compression is a substantially more effective tuning parameter for $T_c$ than in-plane strain, demonstrating that out-of-plane lattice control dominates the competition between superconductivity and the CDW. Furthermore, isolating the cross-coupling coefficient $m_{13}$ reveals a temperature evolution entirely distinct from the effective $A_{1g}$ elastoresistivity measured in standard piezo-based geometries. Specifically, the out-of-plane $A_{1g,2}$ strain contribution, characterized by $m_{13}$, exhibits an order-parameter-like onset across $T_{\mathrm{CDW}}$ and carries an opposite sign relative to the in-plane $A_{1g,1}$ response, $m_{11} + m_{12}$. These results establish the out-of-plane cross-coupling elastoresistivity coefficient $m_{13}$ as a highly sensitive, symmetry-resolved probe of the CDW phase.

\section{Method}

Single crystals of {\CVS} used in this work were grown via the self-flux method following established procedures \cite{ortiz2019new}. The synthesized crystals exhibit a superconducting transition at $T_c \approx 3$ K ($T_{c,\mathrm{offset}} \approx 2.6$ K) and a CDW transition at $T_{\mathrm{CDW}} \approx 94$ K \cite{yang2023thekagome}. After identifying the in-plane crystallographic directions via Laue diffraction, we prepared rectangular bars with their long axes oriented along the crystallographic $a$ axis for in-plane strain measurements, and square platelets for $c$-axis compression measurements, thereby enabling precisely controlled uniaxial loading in both geometries.

Strain-tuning experiments were performed using an FC100 uniaxial stress cell (Razorbill Instruments Ltd.), capable of applying forces up to $F \approx 90$ N at $T=4$ K. For in-plane measurements, thin crystals (typical thickness $\sim 20$ $\mu$m) were bonded to a bowtie-shaped titanium platform to generate precisely controlled in-plane uniaxial strain \cite{park2020rigid}. For out-of-plane measurements, direct $c$-axis compression was applied to a $\sim 90$-$\mu$m-thick, $0.5 \times 0.5$ mm$^2$ square crystal using a titanium-anvil configuration integrated into the FC100 cell, as shown in Figs. 1(e) and 1(f). In both experimental geometries, the longitudinal and transverse resistive responses were measured using a modified Montgomery technique [Fig. 1(c)]. This approach enables the simultaneous determination of the in-plane resistivity components $\rho_{xx}$ ($x \parallel a$) and $\rho_{yy}$ ($y \perp a$) from a single sample, thereby eliminating the cross-contamination of symmetry channels that inherently afflicts differential elastoresistivity measurements \cite{SI}.

\section{Strain tuning of $T_c$ and $T_{\rm CDW}$}

\begin{figure}[htbp!]
\includegraphics[width=7cm]{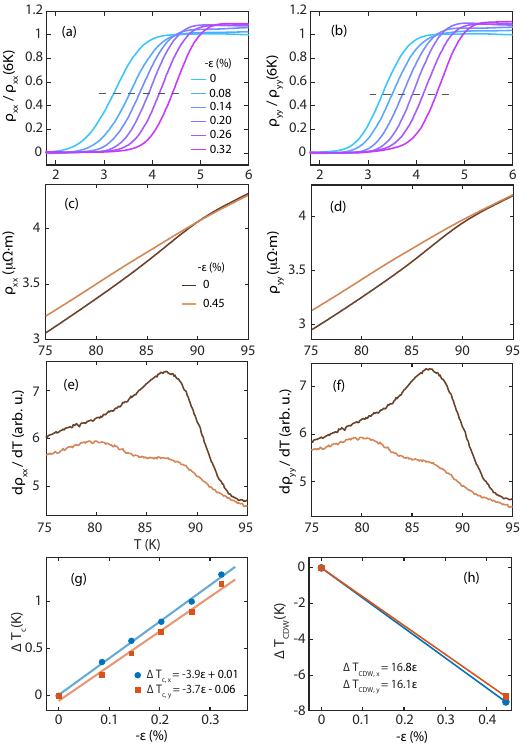}
\caption{$c$-axis compressive strain tuning of $T_c$ and $T_{\rm CDW}$ in {\CVS}. (a,b) Normalized in-plane resistivity $\rho_{xx}(T)/\rho_{xx}(6\,\mathrm{K})$ (a) and $\rho_{yy}(T)/\rho_{yy}(6\,\mathrm{K})$ (b) measured through the superconducting transition under a series of $c$-axis compressive strains. The legend denotes the applied compression plotted as $-\varepsilon$, such that larger $-\varepsilon$ corresponds to stronger compression. Dashed lines indicate the criterion used to determine $\Delta T_c$.
(c,d) Temperature-dependent resistivity $\rho_{xx}(T)$ (c) and $\rho_{yy}(T)$ (d) in the vicinity of the CDW transition under representative $c$-axis compressive strains.
(e,f) Corresponding temperature derivatives $d\rho_{xx}/dT$ (e) and $d\rho_{yy}/dT$ (f), from which $T_{\rm CDW}$ is identified by the peak position. At the largest compression, where two features are resolved, $T_{\rm CDW}$ is extracted from the lower-temperature feature, which shifts systematically with compression and is assigned to the compressed region.
(g) Superconducting transition shift $\Delta T_c$ extracted from panels (a,b) as a function of $-\varepsilon$ for the two current directions; solid lines are linear fits yielding $dT_{c,x}/d\varepsilon$ and $dT_{c,y}/d\varepsilon$.
(h) CDW transition shift $\Delta T_{\rm CDW}$ extracted from panels (e,f) as a function of $-\varepsilon$ for the two current directions; solid lines are linear fits yielding $dT_{{\rm CDW},x}/d\varepsilon$ and $dT_{{\rm CDW},y}/d\varepsilon$.}
\label{fig2}
\end{figure}

The in-plane uniaxial-strain dependence of $T_c$ and $T_{\rm CDW}$ has been established in our previous work \cite{yang2023thekagome} and serves as a benchmark for the present study. For the $c$-axis compression measurements, we employ a cap-block/anvil geometry adapted from established uniaxial-stress designs. To implement this $c$-axis compression, the titanium anvils load a pair of sapphire plates that sandwich the {\CVS} crystal. The sample is bonded to one of the sapphire plates using a thin layer of Stycast 1266 epoxy to ensure uniform stress transfer. The integrated capacitive force sensor of the FC100 cell provides an \textit{in situ} readout of the applied force during voltage sweeps. In practice, a plateau in the capacitance–voltage trace identifies a regime of vanishing applied force, enabling the zero-force point to be determined reproducibly [Fig. 1(g)]. The measured force is converted to uniaxial stress using the manufacturer's calibration curve, while the corresponding uniaxial strain, $\varepsilon_{zz}$, is obtained independently via cryogenic digital image correlation (CDIC) by tracking the relative displacement of fiducial features on the opposing sapphire plates \cite{mo2024cryogenic}. This simultaneous access to both stress and strain allows us to estimate the apparent Young's modulus of the mounted sample configuration [Fig. 1(h)]. After correcting for the finite compliance of the Stycast 1266 layer, we estimate the $c$-axis Young's modulus of {\CVS} to be $Y_c \approx 35\pm3$ GPa, consistent with theoretical calculations of the elastic constants for this material (see the Supplemental Material for details) \cite{SI}.

we estimate the
$c$-axis Young's modulus of CsV$_3$Sb$_5$ to be $Y_c \approx 35 \pm 3$ GPa, consistent with theoretical calculations of the elastic constants for this material (see the Supplemental Material for details) \cite{SI}.

Figure 2 summarizes the evolution of the low-temperature resistive transitions under $c$-axis compression for currents along the $x$ (aligned with the crystallographic $a$ axis) [Figs. 2(a) and 2(c)] and $y$ (the in-plane direction perpendicular to $a$, i.e., $\tfrac{1}{2}\mathbf{a}+\mathbf{b}$) [Figs. 2(b) and 2(d)] directions. In both configurations, $T_c$ increases systematically with increasing compression [Figs. 2(a) and 2(b)], and the enhancement $\Delta T_c$ follows a linear dependence on $\varepsilon_{zz}$ over the full accessible range, with the same slope for $\rho_{xx}$ and $\rho_{yy}$ [Fig. 2(g)]. The magnitude of $dT_c/d\varepsilon_{zz}$ is substantially larger than the corresponding in-plane uniaxial-strain slopes \cite{yang2023thekagome}, underscoring the high efficiency of out-of-plane lattice control. Notably, the linear increase of $T_c$ persists up to the largest $|\varepsilon_{zz}|$ achieved here, suggesting that the enhancement is not yet saturated within our experimental window.

Concomitantly, the CDW transition shifts to lower temperature under c-axis compression [Figs. 2(e), 2(f), and 2(h)]. With increasing $|\varepsilon_{zz}|$, the CDW-related anomaly in $d\rho/dT$ broadens and develops two resolvable features at the largest compression. The higher-temperature feature remains near the zero-strain transition and likely originates from weakly strained or partially relaxed regions, whereas the lower-temperature feature shifts with compression and tracks the compressed region. We therefore extract $T_{\mathrm{CDW}}$ in Fig. 2(h) from the lower-temperature feature. Notably, this possible weakly strained fraction does not affect the extraction of $T_c$, as the superconducting transition remains sharp under c-axis compression, with no resolvable splitting or appreciable broadening. The decrease of $T_{\mathrm{CDW}}$ under compression is consistent with competition between the CDW and superconductivity.

To quantify and separate the in-plane ($A_{1g,1}$) and out-of-plane ($A_{1g,2}$) symmetric contributions to the tuning of $T_c$ and $T_{\mathrm{CDW}}$, we combine the $c$-axis compression results shown in Fig. 2 with the in-plane uniaxial strain data from our previous study \cite{yang2023thekagome}. In our in-plane strain experiments, the sample's transverse in-plane deformation is rigidly constrained by the titanium platform, yielding $\varepsilon_{yy} = -\nu_{12}^{\text{p}}\varepsilon_{xx}$ (where $\nu_{12}^{\text{p}} = 0.32$ is the Poisson ratio of titanium). Meanwhile, the out-of-plane strain is governed by the sample's intrinsic Poisson response, $\varepsilon_{zz} = -\nu_{13}^{\text{s}}\varepsilon_{xx}$. Based on the elastic stiffness constants derived from DFT calculations \cite{frachet2024colossal}, $\nu_{13}^{\text{s}} \approx 0.28$. Therefore, the strain-induced shift $\Delta T_c$ (and analogously for $T_{\mathrm{CDW}}$) under an applied in-plane strain $\varepsilon_{xx}$ can be expressed as a linear combination of the $A_{1g}$ symmetry channels:
\begin{equation}
\Delta T_c(\varepsilon_{xx}) = \left[ \frac{1 - \nu_{12}^{\text{p}}}{2} a_1 - \nu_{13}^{\text{s}} a_2 \right] \varepsilon_{xx} \equiv k_a \varepsilon_{xx},
\end{equation}
where $a_1 \equiv \partial T_c / \partial \varepsilon_{A_{1g,1}}$ and $a_2 \equiv \partial T_c / \partial \varepsilon_{A_{1g,2}}$ represent the pure in-plane and out-of-plane symmetric tuning coefficients, respectively. Conversely, for direct $c$-axis compression, the applied $\varepsilon_{zz}$ induces an isotropic in-plane expansion according to $\varepsilon_{xx} = \varepsilon_{yy} = -\nu_{31}^{\text{s}}\varepsilon_{zz}$, where $\nu_{31}^{\text{s}} \approx 0.10$ reported in \cite{frachet2024colossal}. 
In this nominal free-Poisson limit, the shift under $c$-axis strain is then given by:
\begin{equation}
\Delta T_c(\varepsilon_{zz}) = \left[ -\nu_{31}^{\text{s}} a_1 + a_2 \right] \varepsilon_{zz} \equiv k_c \varepsilon_{zz}.
\end{equation}
Possible partial suppression of the Poisson-induced in-plane expansion by the sapphire/epoxy interface is discussed in the Supplemental Material \cite{SI}. Because the CsV$_3$Sb$_5$ platelet is about $90~\mu$m thick whereas the Stycast 1266 layer is only several micrometers thick and bonded only on one side, this interfacial constraint is not expected to fully clamp the in-plane deformation of the sample. Its influence is included as a systematic uncertainty.
By solving this system of coupled equations using the experimentally measured slopes $k_a$ and $k_c$, we can unambiguously disentangle the $a_1$ and $a_2$ tuning coefficients. The separated $A_{1g,1}$ and $A_{1g,2}$ contributions for both $T_c$ and $T_{\mathrm{CDW}}$ are summarized in Table~I. The statistical and systematic uncertainties associated with these quantities are discussed in the Supplemental Material \cite{SI}.

The decomposition shows that $T_c$ is essentially insensitive to the in-plane symmetric strain, as evidenced by the near-vanishing $a_1$ and its sign change between the $x$ and $y$ configurations, whereas the sizable and nearly identical $a_2$ values in both geometries identify $c$-axis strain as the dominant tuning parameter of $T_{c}$ [Table I]. By contrast, the sizable values of both $a_1$ and $a_2$ for $T_{\mathrm{CDW}}$ indicate that the CDW couples to both $A_{1g,1}$ and $A_{1g,2}$ symmetric channels. This does not contradict previous studies, which inferred from the opposite evolution of $T_{\mathrm{CDW}}$ under hydrostatic pressure and $a$-axis compressive strain that the $c$ axis plays the primary role \cite{qian2021revealing}. Because CsV$_3$Sb$_5$ is elastically much softer along the $c$ axis than in plane, hydrostatic pressure generates a substantially larger $c$-axis strain than in-plane strain, causing the overall response to be dominated by the out-of-plane component. This is illustrated by the hydrostatic-pressure-induced shift of $T_{\mathrm{CDW}}$,
\begin{equation}
\Delta T_{\mathrm{CDW}}(P) = -P\left[\left( \frac{1 - \nu_{12}^{\text{s}}}{Y_a} -\frac{\nu_{31}^{\text{s}}}{Y_c}\right) a_1 + \left(\frac{1}{Y_c}-\frac{2\nu_{13}^{\text{s}}}{Y_a}\right) a_2 \right]  \equiv k_P P,
\end{equation}
where $Y_a \approx 81$ GPa and $Y_c \approx 35$ GPa are the effective Young's moduli along the $a$ and $c$ axes, respectively. Using these values, we obtain $k_P = -0.41 a_1 - 2.17 a_2$, showing that the hydrostatic-pressure response is governed primarily by the much larger $c$-axis strain, which is about five times the in-plane strain and therefore outweighs the stronger intrinsic sensitivity of $T_{\mathrm{CDW}}$ to the latter.

\begin{table}
\caption{\label{tab:pressure_derivatives}
Relative pressure derivatives and symmetric tuning coefficients of the superconducting and CDW transition temperatures in CsV$_3$Sb$_5$.
For hexagonal symmetry, the hydrostatic derivative is
$d\ln T/dp = 2\,d\ln T/dp_a + d\ln T/dp_c$.
}
\begin{ruledtabular}
\begin{tabular}{lcccc}
& $d\ln T/dp_a$ & $d\ln T/dp_c$ & $a_1$ & $a_2$ \\
& (GPa$^{-1}$)  & (GPa$^{-1}$)  & (K/\%) & (K/\%) \\
\hline
$T_c$~~($I\parallel a$)     & $-0.38$ & $+2.94$ & $+0.09$  & $-3.89$ \\
$T_{\rm CDW}$~~($I\parallel a$) & $+0.14$ & $-0.58$ & $-32.13$ & $+13.59$ \\
$T_c$~~($I\perp a$)     & $-0.29$ & $+2.74$ & $-0.21$  & $-3.72$ \\
$T_{\rm CDW}$~~($I\perp a$) & $+0.13$ & $-0.55$ & $-29.46$ & $+13.15$ \\
\hline
\end{tabular}
\end{ruledtabular}
\end{table}

For comparison with the analysis of Frachet \textit{et al.} \cite{frachet2024colossal}, we convert our decomposed strain responses into the relative uniaxial pressure derivatives listed in Table I. For both current configurations, the in-plane and out-of-plane derivatives have opposite signs, and the magnitude of the latter is much larger; for example, $d\ln T_c/dp_c \approx +2.94$ GPa$^{-1}$ is nearly an order of magnitude larger than $d\ln T_c/dp_a \approx -0.38$ GPa$^{-1}$. This comparison further supports the conclusion that the evolution of both $T_c$ and $T_{\mathrm{CDW}}$ is dominated by out-of-plane lattice tuning, consistent with previous experimental and theoretical work emphasizing the exceptional sensitivity of CsV$_3$Sb$_5$ to $c$-axis distortions \cite{qian2021revealing,frachet2024colossal}. At the same time, our experiment accesses this response through direct mechanical $c$-axis compression, rather than through indirect inference based on Poisson-coupled strain, DFT, or thermodynamic analysis.

Although our results confirm the colossal character of the $c$-axis response, the absolute magnitudes of the extracted pressure derivatives are approximately 1.6 to 2 times smaller than those reported in ref. \cite{frachet2024colossal}, where $d\ln T_c/dp_c = +4.7$ GPa$^{-1}$ and $d\ln T_{\mathrm{CDW}}/dp_c = -1.3$ GPa$^{-1}$. This difference likely reflects the distinct methodologies involved. Their values were inferred from thermodynamic relations combined with hydrostatic-pressure data, which rely on specific-heat anomalies, low-strain linearization, and idealized zero-stress assumptions. In contrast, our values are obtained from direct macroscopic stress-strain measurements under $c$-axis loading. In such a geometry, the effective strain transfer can be affected by local strain relaxation, slight anvil misalignment, and the finite stiffness of the epoxy-sapphire mounting stack. Sample-to-sample variations in defect density may also contribute. Despite these quantitative differences, our symmetry decomposition robustly shows that the out-of-plane $A_{1g,2}$ channel is the dominant tuning parameter for both $T_c$ and $T_{\mathrm{CDW}}$.

Beyond clarifying the strain response of {\CVS}, our titanium-anvil setup paired with \textit{in situ} CDIC establishes a robust methodology for tuning electronic phases. By enabling simultaneous control and independent measurement of out-of-plane stress and strain, this direct $c$-axis compression technique offers a highly versatile mechanical probe for van der Waals and quasi-two-dimensional materials, where interlayer spacing often dictates core topological and correlated properties.

\section{$A_{1g,1}$ and $A_{1g,2}$ elastoresistivity}

Elastoresistivity provides a highly sensitive method to access symmetry-resolved electronic susceptibilities by measuring the linear response of the resistivity tensor to controlled strains \cite{kuo2013measurement,kuo2016ubiquitous,nie2022charge}. Standard implementations rely on bonding single crystals to a piezoelectric stack and comparing the longitudinal and transverse resistive responses to isolate specific elastoresistivity coefficients. In the kagome metal {\CVS}, previous works have employed this technique by decomposing the strain and resistivity tensors into the $A_{1g}$ and $E_{2g}$ irreducible representations of the $D_{6h}$ point group \cite{nie2022charge,liu2024absence,frachet2024colossal,asaba2024evidence}. However, the large, active Poisson ratio of typical piezoelectric stacks severely convolutes the in-plane ($A_{1g,1}$) and out-of-plane ($A_{1g,2}$) symmetric strain responses (Eqn. \ref{eq1}). In this work, we deploy our isotropic titanium-platform methodology \cite{park2020rigid,yang2023thekagome} for in-plane measurements ($\nu_{12}^{\text{p}} \approx 0.32$) and exploit direct $c$-axis compression ($\nu_{31}^{\text{s}} \approx 0.1$) to independently access the $A_{1g,2}$ strain channel. This approach allows us to unambiguously disentangle the symmetric elastoresistivity contributions that are inherently mixed in standard experimental geometries.

\begin{figure}[htbp!]
\includegraphics[width=8.5cm]{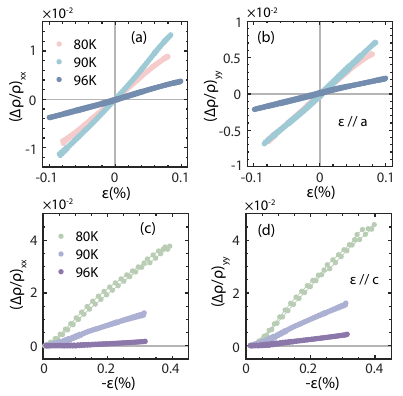}
\caption{Linear strain dependence of the in-plane resistive response under $a$-axis strain and $c$-axis compression.
(a,b) Fractional resistivity change $\Delta\rho/\rho$ measured at selected temperatures (80 K, 90 K, and 96 K) as a function of in-plane uniaxial strain applied along the crystallographic $a$ axis ($\varepsilon \parallel a$). Panel (a) shows the response of $(\Delta\rho/\rho)_{xx}$ (current along $x\parallel a$), while panel (b) shows $(\Delta\rho/\rho)_{yy}$ (current along $y\perp a$).
(c,d) Corresponding $\Delta\rho/\rho$ measured under $c$-axis compression ($\varepsilon\parallel c$), plotted versus $-\varepsilon$ such that increasing values denote increasing compression. Panel (c) shows $(\Delta\rho/\rho)_{xx}$ and panel (d) shows $(\Delta\rho/\rho)_{yy}$.}
\label{fig3}
\end{figure}
 
Figure 3 presents the fractional resistivity changes, $\Delta\rho_{ii}/\rho_{ii}$, at representative temperatures (80 K, 90 K, and 96 K) across the CDW transition. The responses are shown for both in-plane uniaxial strain [Figs. 3(a) and 3(b)] and direct $c$-axis compression [Figs. 3(c) and 3(d)]. In all configurations, the strain-induced resistivity changes remain linear over the accessible strain range, ensuring a robust and well-defined extraction of the underlying elastoresistivity coefficients. As noted above, a weakly strained fraction may be present under c-axis compression, which would mainly dilute the absolute magnitude of $\Delta\rho_{ii}/\rho_{ii}$, rather than change the sign or temperature dependence. Notably, the resistive responses to in-plane strain and $c$-axis strain exhibit opposite signs. For instance, while $\rho_{xx}$ decreases under in-plane compressive strain [Fig. 3(a)], it increases under direct $c$-axis compression [Fig. 3(c)]. This distinct sign reversal naturally mirrors the opposing shifts in $T_c$ and $T_{\mathrm{CDW}}$ observed under these two tuning parameters, physically reflecting their competing effects on the underlying electronic structure and further reinforcing the necessity of separating the $A_{1g,1}$ and $A_{1g,2}$ symmetry channels.

The effective symmetric elastoresistivity measured under in-plane strain on the titanium platform, $m_{A_{1g}}$, provides one linear combination of the pure in-plane and out-of-plane cross-coupling coefficients as shown in Eqn. \ref{eq1}.
Conversely, under direct $c$-axis compression, the primary applied strain $\varepsilon_{zz}$ induces an isotropic in-plane expansion, $\varepsilon_{xx} = \varepsilon_{yy} = -\nu_{31}^{\text{s}}\varepsilon_{zz}$, governed by the sample's Poisson effect.
The effective symmetric elastoresistivity coefficient in this out-of-plane geometry, 
defined as $m_{A_{1g}}^{'} \equiv \zeta_{A_{1g}}/\varepsilon_{zz}$, 
can therefore be expressed as:
\begin{equation}
m_{A_{1g}}^{'} = -\nu_{31}^{\text{s}}(m_{11} + m_{12}) + m_{13}.
\label{eq2}
\end{equation}
Together, Eqns. (1) and (2) form a fully determined linear system. By utilizing the experimentally measured effective slopes ($m_{A_{1g}}$ and $m_{A_{1g}}^{'}$), alongside the known Poisson ratio of the titanium platform ($\nu_{12}^{\text{p}} \approx 0.32$) and the intrinsic sample Poisson ratios ($\nu_{13}^{\text{s}} \approx 0.28$ and $\nu_{31}^{\text{s}} \approx 0.10$) derived from recent theoretical elastic constants \cite{frachet2024colossal}, we can  invert this system.

\begin{figure}[htbp!]
\includegraphics[width=7cm]{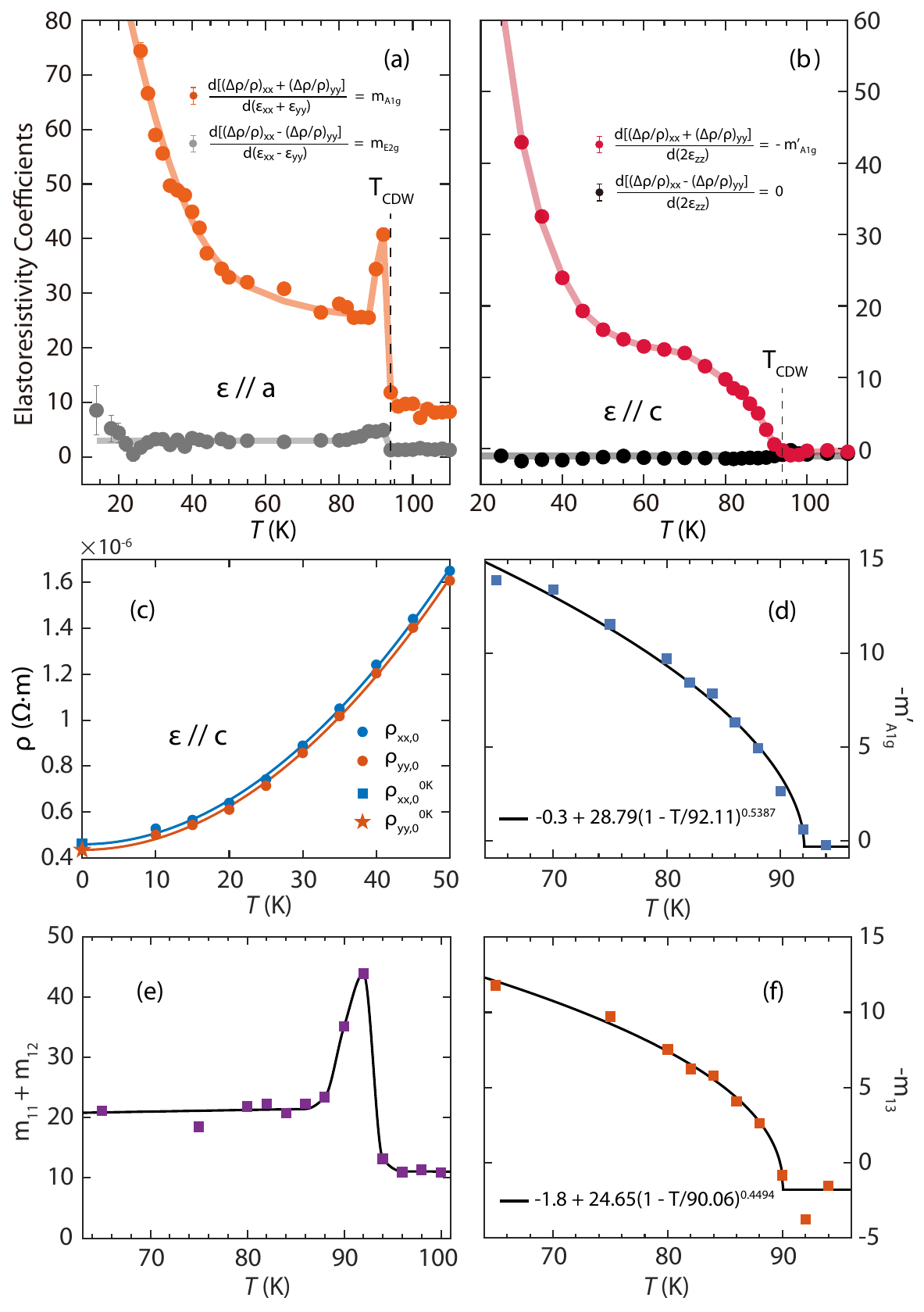}
\caption{Symmetry-resolved elastoresistivity coefficients and isolation of the in-plane and out-of-plane cross-coupling coefficients in {\CVS}. (a) Temperature dependence of the symmetry-resolved elastoresistivity coefficients extracted from in-plane strain sweeps: the anisotropic channel $m_{E_{2g}}$ (gray symbols) and the effective symmetric response $m_{A_{1g}}$ (orange symbols). (b) Temperature dependence of the effective symmetric elastoresistivity coefficient under $c$-axis compression, plotted as $-m'_{A_{1g}}$ (red symbols), together with the difference channel (black symbols), which remains near zero within uncertainty. Vertical lines mark the CDW transition temperature $T_{\rm CDW}$. Solid symbols are obtained after subtracting the strain-independent residual component $\rho_0$ as shown in (c).
(c) Determination of the residual resistivity $\rho_0(T)$ for the $c$-axis compression measurements. The low-temperature resistivity $\rho(T)$ along the $x$ (current $I\parallel a$) and $y$ (current $I\perp a$) directions is fit to $\rho(T)=\rho_0 + AT^2$, yielding $\rho_{xx,0}$ and $\rho_{yy,0}$ (symbols and fits as labeled).
(d) Temperature dependence of the effective out-of-plane symmetric elastoresistivity coefficient $-m'_{A_{1g}}$ just below $T_{\rm CDW}$. The solid line represents a fit capturing the order-parameter-like onset of the response. (e) Extracted pure $A_{1g,1}$ elastoresistivity coefficient, $m_{11} + m_{12}$, obtained by combining the in-plane and $c$-axis data sets in (a) and (b). This successfully isolates the in-plane symmetric response without contamination from the cross-coupling coefficient $m_{13}$.
(f) Extracted pure out-of-plane cross-coupling coefficient $m_{13}$, plotted as $-m_{13}$ to illustrate its opposite sign relative to $m_{11} + m_{12}$. The solid line is a fit to the mean-field form $-m_{13}(T) = -m_{13}^0 + A(1-T/T_{\rm CDW})^{\beta}$ with $\beta \approx 0.5$, capturing the distinct order-parameter-like onset of the $A_{1g,2}$ response below $T_{\rm CDW}$.}
\label{fig4}
\end{figure}

Figure 4 summarizes the temperature dependence of the extracted elastoresistivity coefficients. As shown in Figs. 4(a) and 4(b), both the effective in-plane symmetric elastoresistivity coefficient ($m_{A_{1g}}$) and the effective $c$-axis symmetric elastoresistivity coefficient ($m'_{A_{1g}}$) are small and essentially featureless above the CDW transition. Upon cooling across $T_{\rm CDW}$, both coefficients exhibit dramatic, yet qualitatively distinct, responses. While the in-plane $m_{A_{1g}}$ displays a sharply peaked enhancement consistent with previous reports, the $c$-axis response $m'_{A_{1g}}$ clearly exhibits an order-parameter-like onset. Furthermore, these two symmetry-preserving responses carry opposite signs. We highlight this continuous evolution by fitting the $m'_{A_{1g}}$ data just below the transition [Fig. 4(d)] to a power-law form $\propto (1-T/T_{\rm CDW})^{\beta}$. The fit yields a critical exponent $\beta \approx 0.5$, demonstrating that the out-of-plane symmetric elastoresistivity coefficient couples to the CDW instability in a manner characteristic of a mean-field phase transition.

At lower temperatures (typically below 50 K), the raw elastoresistivity data exhibit sample-dependent upturns or broad maxima. Following the framework established by Frachet \textit{et al.} \cite{frachet2024colossal}, this low-temperature behavior is largely a normalization artifact arising from the strain-insensitive residual resistivity, $\rho_0$. The standard elastoresistivity definition, $m = \frac{1}{\rho}\frac{d\rho}{d\varepsilon}$, mathematically suppresses the intrinsic electronic response as the temperature-dependent scattering rate drops and $\rho$ becomes dominated by the constant $\rho_0$. To access the true thermodynamic strain susceptibility, we fit the low-temperature resistivity to $\rho(T) = \rho_0 + AT^2$ [Fig. 4(c)] and subtract the $\rho_0$ contribution. As plotted with solid symbols in Figs. 4(a) and 4(b), the $\rho_0$-subtracted coefficients ($[1/(\rho-\rho_0)]d\rho/d\varepsilon$) diverge from the raw data, revealing a massive, continuous increase at temperatures below 60 K, consistent with previous studies \cite{frachet2024colossal}. 

Finally, by utilizing the algebraic system defined by Eqns. (1) and (2), we successfully decompose the effective experimental slopes into the pure in-plane symmetric response ($m_{11} + m_{12}$) and the pure out-of-plane cross-coupling coefficient ($m_{13}$). The disentangled components are presented in Figs. 4(e) and 4(f). A direct comparison between the pure in-plane response [Fig. 4(e)] and the raw effective measurement [Fig. 4(a)] reveals that $m_{11} + m_{12}$ is notably smaller in magnitude than $m_{A_{1g}}$. This difference is physically mandated by the opposite signs of the underlying coefficients: because $m_{13}$ is negative, the subtraction of the out-of-plane Poisson contribution ($-m_{13} \frac{2\nu_{ac}}{1-\nu_{ab}}$) artificially inflates the measured $m_{A_{1g}}$ signal. Consequently, conventional in-plane piezo experiments inherently overestimate the pure in-plane symmetric response. The isolated $m_{13}$ coefficient [Fig. 4(f)] proves to be highly robust, maintaining a magnitude comparable to the in-plane response. A mean-field fit to the pure $m_{13}$ data further corroborates the $\beta \approx 0.5$ critical exponent, unambiguously demonstrating that the CDW order parameter in {\CVS} is intimately and fundamentally coupled to out-of-plane symmetric lattice distortions.

\section{Conclusions}
In summary, our combined in-plane uniaxial strain and direct $c$-axis compression measurements on {\CVS} place the response of the intertwined CDW and superconducting states on a rigorously symmetry-resolved footing. We reveal that $c$-axis compression produces a large, linear enhancement of $T_c$ alongside a concomitant suppression of $T_{\rm CDW}$. In addition to acting with an opposite sign, the tuning efficiency of this out-of-plane deformation far exceeds that of in-plane strain, providing direct evidence that $c$-axis lattice control acts as the primary tuning parameter in this material. By utilizing complementary in-plane and $c$-axis elastoresistivity data, we successfully disentangle the fully symmetric responses and isolate $m_{11}+m_{12}$ and $m_{13}$, overcoming the intrinsic cross-channel mixing of $A_{1g,1}$ and $A_{1g,2}$ components inherent to conventional in-plane strain geometries. We find that the out-of-plane cross-coupling coefficient $m_{13}$ is comparable in magnitude but opposite in sign to the pure in-plane response $m_{11}+m_{12}$. Furthermore, $m_{13}$ exhibits a distinct order-parameter-like onset across $T_{\rm CDW}$, fundamentally different from the strongly enhanced, peaked behavior observed in standard in-plane measurements.

Taken together, our results provide direct experimental confirmation that out-of-plane lattice control plays a dominant role in tuning the intertwined CDW and superconducting states in {\CVS}, and establish the out-of-plane cross-coupling elastoresistivity coefficient $m_{13}$ as a highly sensitive, symmetry-resolved probe of the CDW phase. More broadly, the experimental strategy introduced here—combining independent macroscopic strain geometries to unambiguously disentangle symmetric strain channels—offers a general pathway for resolving strain-coupled electronic responses in kagome metals and other layered quantum materials where out-of-plane and in-plane degrees of freedom are strongly intertwined.

\section{Acknowledgments}

\noindent
$\S$ These two authors contributed equally to this work.

The work at BNU is supported by the Scientific Research Innovation Capability Support Project for Young Faculty (ZYGXQNJSKYCXNLZCXM-M2). The work at Zhejiang University was supported by the National Key R\&D Program of China (No. 2022YFA1402200), the National Natural Science Foundation of China (No. 12350710785, and No. 12274363), and the Fundamental Research Funds for the Central Universities (Grant No. 226-2024-00068).


%

\end{document}